\documentstyle[preprint,aps]{revtex}
\tightenlines

\newcommand{\reseteqnum}{\setcounter{equation}{0}}
\newcommand{\nn}{\nonumber}
\newcommand{\eqn}[1]{(\ref{#1})}
\newcommand{\bm}[1]{\mbox{\boldmath $#1$}}
\newcommand{\ovl}[1]{\overline{#1}}
\newcommand{\wt}[1]{\widetilde{#1}}
\newcommand{\wh}[1]{\widehat{#1}}
\newcommand{\p}{\partial}
\newcommand{\dslash}{\p\kern-1.2ex /}
\newcommand{\Dslash}{D\kern-1.5ex /}
\newcommand{\bpsi}{{\overline{\psi}}}
\newcommand{\hpsi}{{\widehat{\psi}}}

\newcommand{\bchi}{{\overline{\chi}}}
\newcommand{\brho}{{\overline{\rho}}}
\newcommand{\bP}{\Pi}
\newcommand{\hP}{{\widehat{\Pi}}}
\newcommand{\ba}{{\overline{a}}}
\newcommand{\bq}{{\overline{q}}}
\newcommand{\tr}{{\rm tr}}

\newcommand{\braket}[2]{\vev{#1 | #2}}
\newcommand{\vev}[1]{\left\langle #1 \right\rangle}

\begin{document}
\draft

\title{
\begin{flushright}
{\normalsize UTHEP-501}\\
{\normalsize UTCCS-P-11}\\
\end{flushright}
Schr\"odinger functional formalism\\
with Ginsparg-Wilson fermion}
\author{
Yusuke Taniguchi}
\address{
Institute of Physics, University of Tsukuba, Tsukuba, Ibaraki 305-8571,
Japan\\}
\date{\today}
\maketitle

\begin{abstract}
 \normalsize
 The Schr\"odinger functional formalism is given as a field theory in a
 finite volume with a Dirichlet boundary condition in temporal
 direction.
 When one tries to construct this formalism with the Ginsparg-Wilson
 fermion including the overlap Dirac operator and the domain-wall
 fermion one easily runs into difficulties.
 The reason is that if the Dirichlet boundary condition is simply
 imposed on the Wilson Dirac operator $D_W$ inside of the overlap Dirac
 operator an exponentially small eigenvalue appears in $D_W$, which
 affects the locality properties of the operator.
 In this paper we propose a new procedure to impose the Schr\"odinger
 functional Dirichlet boundary condition on the overlap Dirac operator
 using an orbifolding projection.
 
\end{abstract}


\newpage
\reseteqnum
\section{Introduction}

The Schr\"odinger functional (SF) is defined as a transition amplitude
between two boundary states with finite time separation
\cite{Symanzik,Luescher85} 
\begin{eqnarray}
Z=\braket{C';x_0=T}{C;x_0=0}=\int{\cal D}\Phi e^{-S[\Phi]}
\end{eqnarray}
and is written in a path integral representation of the field theory
with some specific boundary condition.
\footnote{The same kind of transition amplitude was introduced in
\cite{RossiTesta,Leroy:1990eh} in order to implement the temporal
gauge.}
One of applications of the SF is to define a renormalization scheme
beyond perturbation theory, where the renormalization scale is given by
a finite volume $T\times L^3\sim L^4$ of the system.
The formulation is already accomplished for the non-linear
$\sigma$-model \cite{LWW}, the non-Abelian gauge theory \cite{LNWW} and
the QCD with the Wilson fermion \cite{Sint94,Sint95} including
${\cal O}(a)$ improvement procedure \cite{LSSW9605,LW96}.
(See Ref.~\cite{Sint01} for review.)

In this formalism several renormalization quantities like running gauge
coupling \cite{LSWW93,LSWW94,alpha95,LW95,NW,SS},
Z-factors and ${\cal O}(a)$ improvement factors
\cite{JLLSSSWW,LSSWW,LSSW9611,JS,CLSW}
are extracted conveniently by using a Dirichlet boundary
conditions for spatial component of the gauge field
\begin{eqnarray}
 A_k(x)|_{x_0=0}=C_k(\vec{x}),\quad
 A_k(x)|_{x_0=T}=C_k'(\vec{x})
\label{eqn:DBCgauge}
\end{eqnarray}
and for the quark fields
\begin{eqnarray}
&&
P_+\psi(x)|_{x_0=0}=\rho(\vec{x}),\quad
P_-\psi(x)|_{x_0=T}=\rho'(\vec{x}),
\label{eqn:DBCpsi}
\\&&
\bpsi(x)P_-|_{x_0=0}=\brho(\vec{x}),\quad
\bpsi(x)P_+|_{x_0=T}=\brho'(\vec{x}),
\label{eqn:DBCbpsi}
\\&&
P_\pm=\frac{1\pm\gamma_0}{2}.
\end{eqnarray}
One of privilege of this Dirichlet boundary condition is that the system
acquire a mass gap and there is no infra-red divergence problem.

We notice that the boundary condition is not free to set since it
generally breaks symmetry of the theory and may affect
renormalizability.
However the field theory with Dirichlet boundary condition is shown to
be renormalizable for the pure gauge theory \cite{LNWW}.
And it is also the case for the Wilson fermion \cite{Sint95} by
including a shift in the boundary fields.

Although it is essential to adopt Dirichlet boundary condition for a
mass gap and renormalizability, it has a potential problem of zero mode
in fermion system.
For instance starting from a free Lagrangian
\begin{eqnarray}
{\cal L}=\bpsi\left(\gamma_\mu\p_\mu+m\right)\psi
\end{eqnarray}
with positive mass $m>0$ and the Dirichlet boundary condition
\begin{eqnarray}
P_-\psi|_{x_0=0}=0,\quad
P_+\psi|_{x_0=T}=0
\end{eqnarray}
the zero eigenvalue equation $\left(\gamma_0\p_0+m\right)\psi=0$ in
temporal direction allows a solution
\begin{eqnarray}
\psi=P_+e^{-mx_0}+P_-e^{-m(T-x_0)}
\end{eqnarray}
in $T\to\infty$ limit and a similar solution remains even for finite $T$
with an exponentially small eigenvalue $\propto e^{-mT}$.
In the SF formalism this solution is forbidden by adopting an ``opposite''
Dirichlet boundary condition \eqn{eqn:DBCpsi} and the system has a
finite gap even for $m=0$ \cite{Sint94}.

In the SF formalism of the Wilson fermion \cite{Sint94} we cut the
Wilson Dirac operator at the boundary and the Dirichlet boundary
condition is automatically chosen among
\begin{eqnarray}
P_\pm\psi|_{x_0=0}=0,\quad
P_\mp\psi|_{x_0=T}=0
\end{eqnarray}
depending on signature of the Wilson term.
For example if we adopt a typical signature of the Wilson term
\begin{eqnarray}
D_W=\gamma_\mu\frac{1}{2}\left(\nabla_\mu^*+\nabla_\mu\right)
-\frac{a}{2}\nabla_\mu^*\nabla_\mu+M
\label{eqn:DW}
\end{eqnarray}
the allowed Dirichlet boundary condition is the same as
\eqn{eqn:DBCpsi}.
In this case the zero mode solution can be forbidden by choosing a
proper signature for the mass term; the mass should be kept positive
$M\ge0$ to eliminate the zero mode \cite{Sint94}.

However this zero mode problem may become fatal in the Ginsparg-Wilson
fermion including the overlap Dirac operator
\cite{Neuberger97,Neuberger98} and the domain-wall fermion
\cite{Kaplan,Shamir,FS,KN}.
The overlap Dirac operator is defined by using the Wilson Dirac operator
\eqn{eqn:DW} as
\begin{eqnarray}
D=\frac{1}{\ba}\left(1+D_W\frac{1}{\sqrt{D_W^\dagger D_W}}\right),\quad
\ba=\frac{a}{|M|}.
\label{eqn:OD}
\end{eqnarray}
Here we notice that the Wilson fermion mass $M$ should be kept negative
in a range $-2<M<0$ to impose heavy masses on the doubler modes and a
single massless mode to survive.
As explained in the above when the Dirichlet boundary condition is
imposed directly to the kernel $D_W$ an exponentially small eigenvalue
is allowed for this choice of the Wilson parameter and the mass.
Nearly zero eigenvalue in $D_W$ may break the locality of the overlap
Dirac operator \cite{HJL}.
The situation is also quite similar in the domain-wall fermion.
The same zero mode solution appears in the transfer matrix in fifth
direction, which suppresses the dumping solution in fifth dimension and
allow a chiral symmetry breaking term to appear in the Ward-Takahashi
identity \cite{FS}.

Since a naive formulation of the SF formalism by setting the Dirichlet
boundary condition for the kernel $D_W$ does not work, we need different
procedure to impose boundary condition on the overlap Dirac fermion.
In this paper we propose an orbifolding projection for this purpose.
In section \ref{sec:continuum} we introduce an orbifolding in a
continuum theory and show that the Dirac operator of the orbifolded
theory satisfy the same SF boundary condition.
Although this procedure may be applicable to the general Ginsparg-Wilson
fermion we concentrate on the overlap Dirac fermion in section
\ref{sec:overlap}.
Section \ref{sec:conclusion} is devoted for conclusion.

\reseteqnum
\section{Orbifolding for continuum theory}
\label{sec:continuum}

We notice a fact that the chiral symmetry is broken explicitly by the
Dirichlet boundary condition \eqn{eqn:DBCpsi} in the SF formalism.
This should be also true in the overlap Dirac operator; the
Ginsparg-Wilson relation should be broken in some sense, which was not
accomplished in a naive formulation.
We would adopt this property as a criterion of the SF formalism.

Then we remind a fact that an orbifolded field theory is
equivalent to a field theory with some specific boundary condition.
Since it is possible to break chiral symmetry by an orbifolding
projection in general, it may be able to represent the SF formalism
as an orbifolded theory.
In this section we search for an orbifolding projection which is not
consistent with chiral symmetry and provide the same SF Dirichlet
boundary condition \eqn{eqn:DBCpsi} and \eqn{eqn:DBCbpsi} at fixed
points.

We consider a massless free fermion on $S^1\times\bm{R}^3$
\begin{eqnarray}
{\cal L}=\bpsi(x)\gamma_\mu\p_\mu\psi(x),
\end{eqnarray}
where the anti-periodic boundary condition is set in temporal direction
of length $2T$
\begin{eqnarray}
\psi(\vec{x},x_0+2T)=-\psi(\vec{x},x_0),\quad
\bpsi(\vec{x},x_0+2T)=-\bpsi(\vec{x},x_0).
\label{eqn:APB}
\end{eqnarray}

The orbifolding $S^1/Z_2$ in temporal direction is accomplished by
identifying the negative time with the positive one
$x_0\leftrightarrow-x_0$.
Identification of the fermion field is given by using a symmetry
transformation including the time reflection
\begin{eqnarray}
\psi(x) \to \Sigma\psi(x),\quad
\bpsi(x) \to \bpsi(x)\Sigma,\quad
\Sigma=i\gamma_5\gamma_0R,
\label{eqn:timeref0}
\end{eqnarray}
where $R$ is a time reflection operator
\begin{eqnarray}
R\psi(\vec{x},x_0)=\psi(\vec{x},-x_0).
\end{eqnarray}
$R$ has two fixed points $x_0=0,T$, where $x_0=0$ is a symmetric and
$x_0=T$ is an anti-symmetric fixed point because of the anti-periodicity
\begin{eqnarray}
R\psi(\vec{x},0)=\psi(\vec{x},0),\quad
R\psi(\vec{x},T)=-\psi(\vec{x},T).
\end{eqnarray}
It is free to add any internal symmetry transformation for the
identification and we use the chiral symmetry of the massless fermion
\begin{eqnarray}
\psi(x)\to-i\gamma_5\psi(x),\quad
\bpsi(x)\to-\bpsi(x)i\gamma_5.
\label{eqn:chiraltr}
\end{eqnarray}
Combining \eqn{eqn:timeref0} and \eqn{eqn:chiraltr} we have the
orbifolding symmetry transformation
\begin{eqnarray}
\psi(x) \to -\Gamma\psi(x),\quad
\bpsi(x) \to \bpsi(x)\Gamma,\quad
\Gamma=\gamma_0R.
\label{eqn:orbifoldingtr}
\end{eqnarray}

The orbifolding of the fermion field is given by selecting the following
symmetric sub-space
\begin{eqnarray}
\bP_+\psi(x)=0,\quad
\bpsi(x)\bP_-=0,\quad
\bP_\pm=\frac{1\pm\Gamma}{2}.
\label{eqn:orbifolding-cont}
\end{eqnarray}
We notice that this orbifolding projection provides the proper
homogeneous SF Dirichlet boundary condition at fixed points $x_0=0,T$
\begin{eqnarray}
&&
P_+\psi(x)|_{x_0=0}=0,\quad
P_-\psi(x)|_{x_0=T}=0,
\label{eqn:DBC1}
\\&&
\bpsi(x)P_-|_{x_0=0}=0,\quad
\bpsi(x)P_+|_{x_0=T}=0.
\label{eqn:DBC2}
\end{eqnarray}
The orbifolded action is given by the same projection
\begin{eqnarray}
S=\frac{1}{2}\int d^4x\bpsi(x)D_{\rm SF}\psi(x),\quad
D_{\rm SF}=\bP_+\dslash\bP_-,
\end{eqnarray}
where factor $1/2$ is included since the temporal direction is doubled
compared to the original SF formalism.
We notice that the chiral symmetry is broken explicitly for $D_{\rm SF}$
by the projection.

Now we have two comments.
Since the Schr\"odinger functional of the pure gauge theory is already
well established \cite{LNWW} we treat the gauge field as an external
field and adopt a configuration which is time reflection invariant
\begin{eqnarray}
A_0(\vec{x},-x_0)=-A_0(\vec{x},x_0),\quad
A_i(\vec{x},-x_0)=A_i(\vec{x},x_0)
\end{eqnarray}
and satisfy the SF boundary condition \eqn{eqn:DBCgauge}
simultaneously.
We set periodic boundary condition for the gauge field
\begin{eqnarray}
A_\mu(\vec{x},x_0+2T)=A_\mu(\vec{x},x_0).
\end{eqnarray}
Second comment is on the mass term.
Although the mass term is not consistent with the chiral symmetry,
we can find a symmetric mass term under the orbifolding transformation.
A requirement is that the mass matrix $M$ should anti-commute with
orbifolding operator $\left\{M,\Gamma\right\}=0$.
One of the candidate is a time dependent mass $M=m\eta(x_0)$ with
anti-symmetric and periodic step function
\begin{eqnarray}
&&
\eta(-x_0)=-\eta(x_0),\quad
\eta(x_0+2T)=\eta(x_0),
\nn\\&&
\eta(x_0)=1\quad{\rm for}\quad 0< x_0 <T.
\label{eqn:eta}
\end{eqnarray}

Now the Dirac operator becomes
\begin{eqnarray}
D(m)=\gamma_\mu\left(\p_\mu-iA_\mu(x)\right)+m\eta(x_0),
\end{eqnarray}
which has the orbifolding symmetry
\begin{eqnarray}
\left\{D(m),\Gamma\right\}=0.
\label{eqn:anti-commute}
\end{eqnarray}
The orbifolded action is defined in the same manner as the free theory
\begin{eqnarray}
S=\frac{1}{2}\int d^4x\bpsi(x)D_{\rm SF}(m)\psi(x),\quad
D_{\rm SF}(m)=\bP_+D(m)\bP_-.
\label{eqn:continuum-action}
\end{eqnarray}

We can show that this system with gauge interaction is equivalent to the
original QCD with SF boundary condition \cite{Sint94}.
We consider the bulk region of the temporal direction: $-T<x_0<0$ and
$0<x_0<T$.
In this region the orbifolding condition \eqn{eqn:orbifolding-cont}
becomes
\begin{eqnarray}
&&
\psi(x_0)=-\gamma_0\psi(-x_0),\quad
\bpsi(x_0)=\bpsi(-x_0)\gamma_0
\end{eqnarray}
to identify the fields in positive and negative time.
Combining with the anti-commutative property of the Dirac operator
\eqn{eqn:anti-commute} we can show that contribution to the action from
the bulk region is written as
\begin{eqnarray}
S_{\rm bulk}&=&
\frac{1}{2}\int_{\rm bulk} d^4x \bpsi(x)\Pi_+ D(m) \Pi_-\psi(x)
\nn\\&=&
\int_{0}^Tdx_0 d^3\vec{x}
 \bpsi(x)\left[\gamma_\mu\left(\p_\mu-iA_\mu(x)\right)+m\right]\psi(x),
\end{eqnarray}
which is nothing but the QCD action introduced in Ref.~\cite{Sint94}
except for the surface term, which will be discussed later.
Since the bulk action and the boundary condition for the orbifold
construction is exactly the same as those of QCD with SF boundary
condition, the spectrum and the propagator should be uniquely determined
to be equivalent to those of Ref.~\cite{Sint94} and Ref.~\cite{LW96}.
One can easily check this fact at tree level.

\reseteqnum
\section{Orbifolding for overlap Dirac fermion}
\label{sec:overlap}

Application of the orbifolding procedure is straightforward to the
Ginsparg-Wilson fermions including the overlap Dirac operator
\cite{Neuberger97,Neuberger98}, the domain-wall fermion
\cite{Kaplan,Shamir,FS,KN} and the perfect action
\cite{DHHKN,BW,Hasenfratz} which possess both the time reflection
symmetry
\begin{eqnarray}
\left[\Sigma,D\right]=0
\label{eqn:timeref1}
\end{eqnarray}
and the lattice chiral symmetry \cite{Luescher98} stemming from
the Ginsparg-Wilson relation \cite{GW}
\begin{eqnarray}
&&
\gamma_5D+D\gamma_5=\ba D\gamma_5D.
\label{eqn:GW}
\end{eqnarray}
In this subsection we concentrate on the overlap Dirac operator
\eqn{eqn:OD}, for which the time reflection symmetry \eqn{eqn:timeref1}
comes from that of the Wilson Dirac operator
$\left[\Sigma,D_W\right]=0$.

\subsection{Orbifolding construction of Dirichlet boundary}

As in the continuum case we consider a massless fermion on a
lattice $2N_T\times N_L^3$ with anti-periodic boundary condition in temporal
direction \eqn{eqn:APB}.
We use an orbifolding $S^1/Z_2$ in temporal direction.
Identification of the fermion field is given by using the time
reflection \eqn{eqn:timeref0} and the chiral symmetry of the overlap
Dirac fermion \cite{Luescher98}
\begin{eqnarray}
\psi(x)\to-i\wh{\gamma}_5\psi(x),\quad
\bpsi(x)\to-\bpsi(x)i\gamma_5,\quad
\wh{\gamma}_5=\gamma_5\left(1-\ba D\right),
\label{eqn:chiraltr2}
\end{eqnarray}
where the gauge field is treated as an external field and we adopt a time
reflection symmetric configuration
\begin{eqnarray}
&&
U_k(\vec{x},x_0)=U_k(\vec{x},-x_0),\quad
U_0(\vec{x},x_0)=U_0^\dagger(\vec{x},-x_0-1),
\end{eqnarray}
satisfying the SF Dirichlet boundary condition simultaneously
\begin{eqnarray}
&&
U_k(\vec{x},0)=W_k(\vec{x}),\quad
U_k(\vec{x},N_T)=W_k'(\vec{x}).
\end{eqnarray}

Combining \eqn{eqn:timeref0} and \eqn{eqn:chiraltr2} we have the
orbifolding symmetry transformation
\begin{eqnarray}
\psi(x) \to -\wh{\Gamma}\psi(x),\quad
\bpsi(x) \to \bpsi(x)\Gamma,\quad
\wh{\Gamma}=\Gamma(1-\ba D),
\label{eqn:orbifoldingtr2}
\end{eqnarray}
where $\Gamma$ is the same as the continuum one \eqn{eqn:orbifoldingtr}.
We notice that starting from the time reflection symmetry of the Dirac
operator \eqn{eqn:timeref1} and the Ginsparg-Wilson relation
\eqn{eqn:GW} we have another GW relation for $\Gamma$
\begin{eqnarray}
\Gamma D+D\Gamma=\ba D\Gamma D
\end{eqnarray}
and $\Gamma$ Hermiticity
\begin{eqnarray}
\Gamma D\Gamma=D^\dagger.
\end{eqnarray}
The operator $\wh{\Gamma}$ has a property $\wh{\Gamma}^2=1$ like
$\Gamma$ and can be used to define a projection operator in the
following.

The orbifolding identification of the fermion field is given in the same
way with slightly different projection operator
\begin{eqnarray}
\hP_+\psi(x)=0,\quad
\bpsi(x)\bP_-=0,\quad
\hP_\pm=\frac{1\pm\wh{\Gamma}}{2},
\label{eqn:orbifolding}
\end{eqnarray}
which turn out to be the SF Dirichlet boundary condition \eqn{eqn:DBC1}
and \eqn{eqn:DBC2} at fixed points in the continuum limit.
Using the time reflection symmetry \eqn{eqn:timeref1} we can easily show
that the projection operators $\Gamma$ and $\wh{\Gamma}$ do not have an
``index''\footnote{
The same property is satisfied for the chiral index
$\tr\wh{\gamma}_5=\tr(\wh{\gamma}_5\Sigma^2)=-\tr\wh{\gamma}_5=0$.
}
\begin{eqnarray}
\tr\Gamma=\tr\hat{\Gamma}=0
\end{eqnarray}
and furthermore we can find a local unitary transformation
\begin{eqnarray}
u=\frac{1+\Sigma}{2}(1-\ovl{a}D)+\frac{1-\Sigma}{2},\quad
u'=\gamma_5u\gamma_5,
\label{eqn:unitary}
\end{eqnarray}
which connects $\wh{\Gamma}$ and $\Gamma$ as
\begin{eqnarray}
\wh{\Gamma}=u^\dagger\Gamma u,\quad
\wh{\Gamma}=u'\Gamma u'^\dagger.
\end{eqnarray}
The projection operator $\hP_\pm$ spans essentially the same Hilbert
sub-space as $\bP_\pm$.
We notice that this unitary operator connects $\wh{\gamma}_5$ and
$\gamma_5$ in a similar way
\begin{eqnarray}
\wh{\gamma}_5=u^\dagger\gamma_5 u,\quad
\wh{\gamma}_5=u'\gamma_5 u'^\dagger.
\end{eqnarray}

The physical quark operator is defined to transform in a same manner as
the continuum under chiral rotation,
\begin{eqnarray}
\delta q(x)=\gamma_5q(x),\quad
\delta\bq(x)=\bq(x)\gamma_5.
\end{eqnarray}
Since we have a unitary operator $u$ and $u'$ we have several ways 
to define a physical quark field from GW fermion fields $\psi$ and
$\bpsi$.\footnote{
One may suppose that a local Dirac operator can be constructed with
exact chiral symmetry by adopting the physical quark operator of
\eqn{eqn:quark2} and \eqn{eqn:quark3}, where the effective action of
$q$, $\bq$ is kept local.
However the unitary transformation $u$ contains the time reflection
operator $R$ and furthermore the gauge configuration is fixed to be time
reflection invariant.
There is no translation invariance in the physical quark effective
theory and this is consistent with the Nielsen-Ninomiya no-go theorem.
}
For example
\begin{eqnarray}
&&
q(x)=\left(1-\frac{\ba}{2}D\right)\psi(x),\quad
\bq(x)=\bpsi(x),
\label{eqn:quark1}
\\&&
q(x)=u\psi(x),\quad
\bq(x)=\bpsi(x),
\label{eqn:quark2}
\\&&
q(x)=u'^\dagger\psi(x),\quad
\bq(x)=\bpsi(x).
\label{eqn:quark3}
\end{eqnarray}
These three definitions are not independent but connected with
$u+u'^\dagger=\left(2-\ba D\right)$.
The orbifolding of the physical quark field becomes the same as that of
the continuum theory
\begin{eqnarray}
\bP_+q(x)=0,\quad
\bq(x)\bP_-=0.
\end{eqnarray}

The massless orbifolded action is given by
\begin{eqnarray}
S=\frac{1}{2}a^4\sum\bpsi D_{\rm SF}\psi,\quad
D_{\rm SF}=\bP_+D\hP_-.
\label{eqn:massless}
\end{eqnarray}
We have four comments here.
(i) It should be emphasized that the SF Dirac operator
$D_{\rm SF}$ is local since it is constructed by multiplying local
objects only.
(ii) The massless SF Dirac operator $D_{\rm SF}$ does not satisfy the
chiral Ginsparg-Wilson relation \eqn{eqn:GW} and the chiral symmetry is
broken by projection as was expected.
(iii) Although two different projection operators $\Gamma$ and
$\wh{\Gamma}$ are used from the left and right of $D_{\rm SF}$ this does
not bring the problem we encountered in the chiral gauge theory since
these two operators are connected by the unitary transformation $u$ or
$u'$.
(iv) Since we have some ambiguity in defining the mass term we
concentrate on the massless case and postpone the massive theory to the
later sub-section.

It is clear that \eqn{eqn:massless} is a well defined lattice
regularization of the SF QCD with gauge interaction
since the Dirac operator is local and
the lattice action \eqn{eqn:massless} converges to the continuum
one \eqn{eqn:continuum-action} in $a\to0$ limit including the boundary
condition.
We shall see explicit scaling behavior for free theory in the following
two sub-sections.

\subsection{Eigenvalue of free Dirac operator}

We consider an eigenvalue problem of the massless SF Dirac operator
\begin{eqnarray}
D_{\rm SF}=\bP_+D\hP_-
\end{eqnarray}
at tree level.
This Dirac operator connects two different Hilbert sub-space as
in the continuum \cite{Sint94}
\begin{eqnarray}
&&
D_{\rm SF} : \wh{\cal H}_+\to{\cal H}_-,\quad
D_{\rm SF}^\dagger : {\cal H}_-\to\wh{\cal H}_+,
\label{eqn:subspace}
\\&&
\wh{\cal H}_+=\left\{\psi|\hP_+\psi=0\right\},\quad
{\cal H}_-=\left\{\psi|\Pi_-\psi=0\right\}.
\end{eqnarray}
We need to introduce the ``doubled'' Hermitian Dirac operator
\begin{eqnarray}
{\cal D}=\pmatrix{&D_{\rm SF}\cr D_{\rm SF}^\dagger\cr},
\label{eqn:doubledD}
\end{eqnarray}
which connects the same Hilbert space
${\cal H}_-\oplus\wh{\cal H}_+\to{\cal H}_-\oplus\wh{\cal H}_+$.
As in the continuum the eigenvalue problem is solved on a two component
vector
\begin{eqnarray}
\Psi=\pmatrix{\psi_-\cr\hpsi_+\cr},\quad
\psi_-\in{\cal H}_-,\quad
\hpsi_+\in\wh{\cal H}_+
\label{eqn:vector}
\end{eqnarray}
and the eigenvalue equation is written in a following form
\begin{eqnarray}
D_{\rm SF}\hpsi_+=\lambda\psi_-,\quad
D_{\rm SF}^\dagger\psi_-=\lambda\hpsi_+
\label{eqn:ev}
\end{eqnarray}
with a real eigenvalue $\lambda$.

A candidate of the eigen-function in each Hilbert sub-space
$\wh{\cal H}_+$ and ${\cal H}_-$ is given by 
\begin{eqnarray}
\wh{\psi}_+(x)=u^\dagger\psi_+(x),\quad
\psi_+(x)= f_+(x_0)ve^{i\vec{p}\vec{x}},\quad
\psi_-(x)=f_-(x_0)we^{i\vec{p}\vec{x}},
\label{eqn:eigenfunction}
\end{eqnarray}
where $f_\pm$ are given to satisfy $\Gamma f_\pm=\mp f_\pm$ as
\begin{eqnarray}
&&
f_+(x_0)=\alpha\left(P_+\sin p_0x_0+P_-\cos p_0x_0\right),
\\&&
f_-(x_0)=\alpha\left(P_+\cos p_0x_0+P_-\sin p_0x_0\right)
\end{eqnarray}
with
\begin{eqnarray}
p_0=\frac{2n-1}{2N_T}\pi,\quad
n=\left\{-N_T+1,\cdots,N_T\right\}
\label{eqn:p0}
\end{eqnarray}
for anti-periodicity in $2N_T$.
$v$ and $w$ are four component vectors in spinor space.
In the following we consider the operator
\begin{eqnarray}
D_{\rm SF}u^\dagger=
\Pi_+\left\{
\frac{1}{2}\left(D-D^\dagger\right)-\Sigma\frac{1}{2}\left(D+D^\dagger\right)
\right\}\Pi_-
\end{eqnarray}
and its operation on $\psi_+$.

We start from the fact that $\psi_+$ is an eigen-function of the
operator $D_W^\dagger D_W$ and $D_W+D_W^\dagger$ with eigenvalues
$\lambda_W^2$ and $2W$ where
\begin{eqnarray}
\lambda_W^2=\sum_{\mu}\sin^2ap_\mu+W^2,\quad
W=-|M|+\sum_{\mu}\left(1-\cos ap_\mu\right).
\end{eqnarray}
On the other hand $D_W-D_W^\dagger$ operates as
\begin{eqnarray}
\frac{1}{2}\left(D_W-D_W^\dagger\right)\psi_+(x)&=&
\left(\sin ap_0f_-(x_0)+i\gamma_i\sin ap_if_+(x_0)\right)ve^{i\vec{p}\vec{x}}.
\end{eqnarray}
Combining these two and taking similar manipulation for
$uD_{\rm SF}^\dagger$ on $\psi_-$ the eigenvalue equation is re-written
as
\begin{eqnarray}
&&
D_{\rm SF}\hpsi_+(x)
=f_-(x_0)\left(A_0+i\gamma_iA_i+i\gamma_5B\right)ve^{i\vec{p}\vec{x}}
=\lambda\psi_-(x),\quad
\\&&
D_{\rm SF}^\dagger\psi_-(x)
=u^\dagger f_+(x_0)
 \left(A_0-i\gamma_iA_i-i\gamma_5B\right)we^{i\vec{p}\vec{x}}
=\lambda\hpsi_+(x)
\end{eqnarray}
with
\begin{eqnarray}
&&
A_\mu=\frac{\sin ap_\mu}{\ba\sqrt{\lambda_W^2}},\quad
B=\frac{1}{\ba}\left(1+\frac{W}{\sqrt{\lambda_W^2}}\right).
\label{eqn:AB}
\end{eqnarray}
Here we used $\Sigma=i\gamma_5\Gamma$ and a property
$\Gamma f_\pm=\mp f_\pm$.
At last we solve the eigenvalue equation in the spinor space and get the
eigenvalue
\begin{eqnarray}
\lambda^2=A_\mu^2+B^2.
\label{eqn:EV}
\end{eqnarray}
This eigenvalue agrees with that of the massless continuum theory
\cite{Sint94} in $a\to0$ limit combined with the discretization
condition \eqn{eqn:p0}.

\subsection{Free propagator}

The fermion propagator is formally given by the inverse of the Dirac
operator
\begin{eqnarray}
G_{\rm SF}(x,y)
=2\left(D_{\rm SF}^{-1}\right)_{x,y}
=2\left(\hP_-\frac{1}{D}\bP_+\right)_{x,y}.
\label{eqn:prop}
\end{eqnarray}
where inverse is defined in the Hilbert sub-space
$\wh{\cal H}_+$ or ${\cal H}_-$ as
\begin{eqnarray}
D_{\rm SF}D_{\rm SF}^{-1}=\Pi_+,\quad
D_{\rm SF}^{-1}D_{\rm SF}=\hP_-.
\end{eqnarray}
At tree level this propagator can be written in a simple form as
\begin{eqnarray}
&&
G_{\rm SF}=2D^\dagger\bP_+\frac{1}{DD^\dagger}\bP_+
=D^\dagger\left(P_+G_L+P_-G_R\right),
\\&&
G_{\rm R/L}(x,y)=
\frac{1}{N_L^3}\sum_{\vec{p}}e^{i\vec{p}\left(\vec{x}-\vec{y}\right)}
G_{\rm R/L}(x_0,y_0;\vec{p}),
\\&&
G_{\rm R}(x_0,y_0;\vec{p})=\frac{1}{2aN_T}\sum_{n=-N_T+1}^{N_T}
\frac{1}{A_\mu^2+B^2}\left(e^{ip_0(x_0-y_0)}-e^{ip_0(x_0+y_0)}\right),
\\&&
G_{\rm L}(x_0,y_0;\vec{p})=\frac{1}{2aN_T}\sum_{n=-N_T+1}^{N_T}
\frac{1}{A_\mu^2+B^2}\left(e^{ip_0(x_0-y_0)}+e^{ip_0(x_0+y_0)}\right),
\end{eqnarray}
where $p_0$ satisfy the quantization condition \eqn{eqn:p0}.

If we adopt the definition \eqn{eqn:quark1} for the physical quark field
its propagator is given by
\begin{eqnarray}
\vev{q(x)\bq(y)}=\left[\left(1-\frac{\ovl{a}}{2}D\right)G_{\rm SF}\right](x,y).
\end{eqnarray}
The proper Dirichlet boundary condition \cite{LW96} is trivially
satisfied for this quark propagator
\begin{eqnarray}
&&
P_+\vev{q(x)\bq(y)}|_{x_0=0}=0,\quad
P_-\vev{q(x)\bq(y)}|_{x_0=T}=0,
\\&&
\vev{q(x)\bq(y)}|_{y_0=0}P_-=0,\quad
\vev{q(x)\bq(y)}|_{y_0=T}P_+=0
\end{eqnarray}
because of the projection $\Pi_\pm$.
At tree level the propagator takes the form
\begin{eqnarray}
a^3\sum_{\vec{x}}e^{-i\vec{p}\left(\vec{x}-\vec{y}\right)}\vev{q(x)\bq(y)}&=&
\frac{1}{2aN_T}\sum_{n=-N_T+1}^{N_T}
\left(\frac{-i\gamma_\mu A_\mu+B}{A_\mu^2+B^2}-\frac{\ovl{a}}{2}\right)
\nn\\&&\times
e^{ip_0x_0}
\left\{\left(e^{-ip_0y_0}+e^{ip_0y_0}\right)P_+
+\left(e^{-ip_0y_0}-e^{ip_0y_0}\right)P_-\right\},
\end{eqnarray}
which can be shown to approach to the continuum SF propagator
\cite{LW96} without any ${\cal O}(a)$ term.

\subsection{Phase of Dirac determinant}

In general the determinant of the SF Dirac operator is not real for the
overlap fermion since there is no $\gamma_5$ Hermiticity.
Instead we have a following ``Hermiticity'' relation
\begin{eqnarray}
\gamma_5uD_{\rm SF}u^\dagger\gamma_5=D_{\rm SF}^\dagger,\quad
\gamma_5u'^\dagger D_{\rm SF}u'\gamma_5=D_{\rm SF}^\dagger.
\label{eqn:hermiticity}
\end{eqnarray}
However one cannot conclude reality from this relation since the SF
Dirac operator connects different Hilbert sub-space as in
\eqn{eqn:subspace} and the determinant cannot be evaluated directly with
$D_{\rm SF}$.
We need to make a ``Hermitian'' Dirac operator which connects the
same Hilbert sub-space in order to define the Dirac determinant.
This is accomplished by $u^\dagger\gamma_5$ or $u'\gamma_5$, which
turns out to be $\gamma_5$ in the continuum.
We define
\begin{eqnarray}
&&
H_{\rm SF}=D_{\rm SF}u^\dagger\gamma_5=\bP_+Du^\dagger\gamma_5\bP_+
\quad:\quad{\cal H}_-\to{\cal H}_-,
\label{eqnhermitian1}
\\&&
H_{\rm SF}'=D_{\rm SF}u'\gamma_5=\bP_+Du'\gamma_5\bP_+
\quad:\quad{\cal H}_-\to{\cal H}_-.
\label{eqnhermitian2}
\end{eqnarray}
The determinant is evaluated on the sub-space ${\cal H}_-$
\begin{eqnarray}
&&
\det_{\{{\cal H_-}\}}H_{\rm SF}
=\det\left(\bP_+Du^\dagger\gamma_5\bP_++\bP_-\right),
\\&&
\det_{\{{\cal H_-}\}}H_{\rm SF}'
=\det\left(\bP_+Du'\gamma_5\bP_++\bP_-\right),
\end{eqnarray}
where the right hand side is understood to be evaluated in the full
Hilbert space by filling the opposite sub-space ${\cal H}_+$ with
unity.

The phase of the determinant is given as follows
\begin{eqnarray}
&&
\left(\det_{\{{\cal H_-}\}}H_{\rm SF}^{(\prime)}\right)^*=
e^{-2i\phi^{(\prime)}}\left(\det_{\{{\cal H_-}\}}H_{\rm SF}^{(\prime)}\right),
\\&&
e^{-2i\phi}=\det_{\{{\cal H_-}\}}\left(\gamma_5u\right)^2=\det u,
\label{eqn:phase1}
\\&&
e^{-2i\phi'}=\det_{\{{\cal H_-}\}}\left(\gamma_5u'^\dagger\right)^2
=\det u^\dagger=e^{2i\phi},
\label{eqn:phase2}
\end{eqnarray}
which is not real in general.
In the second equality of \eqn{eqn:phase1} and \eqn{eqn:phase2} we used
a relation
\begin{eqnarray}
\Sigma=\omega\Gamma\omega^\dagger,\quad
\Gamma=\omega^\dagger\Sigma\omega,\quad
\omega=e^{i\frac{\pi}{4}\gamma_5}.
\end{eqnarray}
The determinant of the unitary operator $u$ is given by a product of
eigenvalues $\lambda_n$ of the overlap Dirac operator 
\begin{eqnarray}
\det u=\prod_{n\in\{+\}}(1-a\lambda_n),
\end{eqnarray}
where product is taken over a sub-space in which the eigenvalue of
$\Sigma=+1$ and the conjugate eigenvalue $\lambda_n^*$ does not
necessarily belongs to this sub-space.

However we notice that this complexity of the Dirac determinant is not an
essential problem since the phase is an ${\cal O}(a)$ irrelevant effect
and disappears in the continuum limit.
Furthermore if we consider variation of the phase
\begin{eqnarray}
\delta_{\epsilon(x)}\phi=\frac{i}{2}\tr\delta_{\epsilon(x)}uu^{-1}
=-\frac{i}{4}a\tr\left[\Sigma\delta_{\epsilon(x)}D\left(1-aD^\dagger\right)
\right]
\label{eqn:phasevar}
\end{eqnarray}
under a local variation of the link variable
\begin{eqnarray}
\delta_{\epsilon(x)}U_\mu(x)=a\epsilon_\mu(x)U_\mu(x)
\end{eqnarray}
we can show that $\delta_{\epsilon(x)}\phi$ is localized at the
boundary.
Since $\Sigma$ contains time reflection $R$ and both of the operator
$\delta D$ and $\left(1-aD^\dagger\right)$ are local, the trace in
\eqn{eqn:phasevar} has a contribution only at the boundary.
Contribution from the bulk is suppressed exponentially by the locality
property.

For practical application to numerical simulation this phase problem
should be settled.
This is possible for even number of flavours by using two definitions of
``Hermitian'' Dirac operator \eqn{eqnhermitian1} \eqn{eqnhermitian2}
and a fact that the phase is opposite for these definitions as shown in
\eqn{eqn:phase1} \eqn{eqn:phase2}.
The phase can be absorbed into re-definition of the fermion fields.
Explicit form of the two flavour Hermitian Dirac operator is
\begin{eqnarray}
H_{\rm SF}^{(2)}
=\pmatrix{D_{\rm SF}\cr&D_{\rm SF}\cr}U^\dagger\gamma_5\tau^{1,2}
\quad:\quad{\cal H}_-\oplus{\cal H}_-\to{\cal H}_-\oplus{\cal H}_-,
\label{eqn:hermite0}
\end{eqnarray}
where the unitary matrix is defined as
\begin{eqnarray}
U=\pmatrix{u\cr&u'^\dagger\cr}
\label{eqn:unitary2}
\end{eqnarray}
to act on flavour space and $\tau^a$ is the Pauli matrix.
$\gamma_5$ is also understood as two by two on flavour space.
The Dirac operator is exactly Hermite for this definition:
$H_{\rm SF}^{(2)\dagger}=H_{\rm SF}^{(2)}$ and the determinant is real.

For single flavour case it is not still clear how to solve this
practical problem.
However the determinant is real at tree level and one may expect that
the phase is settled as one approaches to the continuum limit.

\subsection{Surface term}

When extracting the renormalization factors of fermions it is
convenient to consider a operator involving the boundary source fields
\begin{eqnarray}
&&
\zeta(\vec{x})=\frac{\delta}{\delta\ovl{\rho}(\vec{x})},\quad
\ovl{\zeta}(\vec{x})=-\frac{\delta}{\delta\rho(\vec{x})},
\\&&
\zeta'(\vec{x})=\frac{\delta}{\delta\ovl{\rho}'(\vec{x})},\quad
\ovl{\zeta}'(\vec{x})=-\frac{\delta}{\delta\rho'(\vec{x})},
\end{eqnarray}
where $\rho, \cdots, \ovl{\rho}'$ are boundary values of the fermion
fields given in \eqn{eqn:DBCpsi} and \eqn{eqn:DBCbpsi}.
Coupling of the boundary value to the bulk dynamical fields was
naturally introduced in the Wilson fermion \cite{Sint94}.
However this is not the case for our construction since the boundary
value vanishes with the orbifolding projection.

In this paper we regard the boundary value as an external source field
and introduce its coupling with the bulk fields according to the
criteria: the coupling terms (surface terms) are local and reproduce the
same form of the correlation function between the boundary fields in the
continuum limit.
Here we define the boundary values on the physical quark fields
\begin{eqnarray}
&&
P_+q(x)|_{x_0=0}=\rho(\vec{x}),\quad
P_-q(x)|_{x_0=N_T}=\rho'(\vec{x}),
\label{eqn:DBCq}
\\&&
\bq(x)P_-|_{x_0=0}=\brho(\vec{x}),\quad
\bq(x)P_+|_{x_0=N_T}=\brho'(\vec{x}).
\label{eqn:DBCbq}
\end{eqnarray}
One of candidates of the surface term is
\begin{eqnarray}
S_{\rm surface}&=&a^3\sum_{\vec{x}}\Bigl(
-\left.\ovl{\rho}(\vec{x})P_-q(x)\right|_{x_0=0}
-\left.\bq(x)P_+\rho(\vec{x})\right|_{x_0=0}
\nn\\&&
-\left.\ovl{\rho}'(\vec{x})P_+q(x)\right|_{x_0=N_T}
-\left.\bq(x)P_-\rho'(\vec{x})\right|_{x_0=N_T}
\Bigr),
\label{eqn:surface}
\end{eqnarray}
where $q$ and $\bq$ are active dynamical fields on the boundary.

According to Ref.~\cite{LW96} we introduce the generating functional
\begin{eqnarray}
Z_F\left[\ovl{\rho}',\rho';\ovl{\rho},\rho;\ovl{\eta},\eta;U\right]&=&
\int D\psi D\bpsi \exp\biggl\{
-S_F\left[U,\bpsi,\psi;\ovl{\rho}',\rho',\ovl{\rho},\rho\right]
\nn\\&&
+a^4\sum_{x}\left(\bpsi(x)\eta(x)+\ovl{\eta}(x)\psi(x)\right)
\biggr\},
\end{eqnarray}
where $\eta(x)$ and $\ovl{\eta}(x)$ are source fields for the fermion
fields and the total action $S_F$ is given as a sum of the bulk action
\eqn{eqn:massless} and the surface term \eqn{eqn:surface}.
We notice that the fermion fields $\psi$ and $\bpsi$ obey the
orbifolding condition \eqn{eqn:orbifolding}.
We decompose the fermion fields into classical and quantum components
\begin{eqnarray}
\psi(x)=\psi_{\rm cl}(x)+\chi(x),\quad
\bpsi(x)=\bpsi_{\rm cl}(x)+\bchi(x)
\end{eqnarray}
and insert it into the generating functional.
The correlation functions between the boundary fields are derived with
the same procedure as Ref.~\cite{LW96} by making use of this
decomposition.
\begin{eqnarray}
&&
\vev{\psi(x)\bpsi(y)}=G_{\rm SF}(x,y),
\\&&
\vev{q(x)\bq(y)}=\left[\left(1-\frac{\ovl{a}}{2}D\right)G_{\rm SF}\right](x,y),
\\&&
\vev{q(x)\ovl{\zeta}(\vec{y})}=
\left.\left[\left(1-\frac{\ovl{a}}{2}D\right)G_{\rm SF}\right](x,y)
P_+\right|_{y_0=0},
\\&&
\vev{q(x)\ovl{\zeta}'(\vec{y})}=
\left.\left[\left(1-\frac{\ovl{a}}{2}D\right)G_{\rm SF}\right](x,y)
P_-\right|_{y_0=N_T},
\\&&
\vev{\zeta(\vec{x})\bq(y)}=
P_-\left.\left[\left(1-\frac{\ovl{a}}{2}D\right)G_{\rm SF}\right](x,y)
\right|_{x_0=0},
\\&&
\vev{\zeta'(\vec{x})\bq(y)}=
P_+\left.\left[\left(1-\frac{\ovl{a}}{2}D\right)G_{\rm SF}\right](x,y)
\right|_{x_0=N_T},
\\&&
\vev{\zeta(\vec{x})\ovl{\zeta}(\vec{y})}=
P_-\left.\left[\left(1-\frac{\ovl{a}}{2}D\right)G_{\rm SF}\right](x,y)P_+
\right|_{x_0=0,y_0=0},
\\&&
\vev{\zeta(\vec{x})\ovl{\zeta}'(\vec{y})}=
P_-\left.\left[\left(1-\frac{\ovl{a}}{2}D\right)G_{\rm SF}\right](x,y)P_-
\right|_{x_0=0,y_0=N_T},
\\&&
\vev{\zeta'(\vec{x})\ovl{\zeta}(\vec{y})}=
P_+\left.\left[\left(1-\frac{\ovl{a}}{2}D\right)G_{\rm SF}\right](x,y)P_+
\right|_{x_0=N_T,y_0=0},
\\&&
\vev{\zeta'(\vec{x})\ovl{\zeta}'(\vec{y})}=
P_+\left.\left[\left(1-\frac{\ovl{a}}{2}D\right)G_{\rm SF}\right](x,y)P_-
\right|_{x_0=N_T,y_0=N_T}.
\end{eqnarray}
Here we adopted the physical quark field of \eqn{eqn:quark1}.
The propagator $G_{\rm SF}$ is defined in \eqn{eqn:prop}.
We notice that the above propagators between the boundary fields and
physical quark fields approach to the continuum SF boundary propagator
without any ${\cal O}(a)$ term.

\subsection{Mass term}

The mass term may be introduced with the same procedure as the continuum
theory.
We consider a mass matrix $M$ which is consistent with the orbifolding
symmetry
\begin{eqnarray}
\Gamma M+M\wh{\Gamma}=0.
\label{eqn:symmass}
\end{eqnarray}
Since the orbifolding transformation is the same as the continuum one on
the physical quark fields, a naive candidate is to couple the continuum
mass matrix $m\eta(x_0)$ to the physical scalar density consisting of
$q(x)$ and $\bq(x)$.
Corresponding to various definition of the quark fields
\eqn{eqn:quark1}-\eqn{eqn:quark3} we have several definitions of the
mass term
\begin{eqnarray}
{\cal L}_m=
m\bpsi\eta\left(1-\frac{\ovl{a}}{2}D\right)\psi,\quad
m\bpsi\eta u\psi,\quad
m\bpsi\eta u'^\dagger\psi,
\end{eqnarray}
where $\eta$ is an anti-symmetric step function \eqn{eqn:eta} on
lattice.

However we encounter a problem with this naive definition of mass term,
since the massive Dirac operator does not satisfy the ``Hermiticity''
relation \eqn{eqn:hermiticity}.
The phase of the Dirac determinant becomes mass dependent although it is
still irrelevant ${\cal O}(a)$ term.

In order to avoid this unpleasant situation we propose even flavors
formulation.
For two flavors case we define the two by tow Dirac operator as
\begin{eqnarray}
D_{\rm SF}^{(2)}(m)=\pmatrix{D_{\rm SF}(m)_1\cr&D_{\rm SF}(m)_2},
\end{eqnarray}
where
\begin{eqnarray}
&&
D_{\rm SF}(m)_1=\bP_+
\left(D+m\eta\left(1-\frac{\ovl{a}}{2}D\right)\right)\hP_-
\\&&
D_{\rm SF}(m)_2
=\bP_+\left(D+m\left(1-\frac{\ovl{a}}{2}D\right)u'\eta u'^\dagger\right)
\hP_-.
\end{eqnarray}
A ``Hermiticity'' relation can be found for this two flavors Dirac
operator as
\begin{eqnarray}
D_{\rm SF}^{(2)}(m)^\dagger=
 \tau^{1,2}\gamma_5UD_{\rm SF}^{(2)}(m)U^\dagger\gamma_5\tau^{1,2},
\label{eqn:hermiticity2}
\end{eqnarray}
where $U$ is defined in \eqn{eqn:unitary2}.

The Hermitian Dirac operator can be defined to connect the same Hilbert
sub-space as
\begin{eqnarray}
H_{\rm SF}^{(2)}(m)=D_{\rm SF}^{(2)}(m)U^\dagger\gamma_5\tau^{1,2}
\quad:\quad{\cal H}_-\oplus{\cal H}_-\to{\cal H}_-\oplus{\cal H}_-,
\label{eqn:hermitem}
\end{eqnarray}
which is re-written in a trivially Hermitian form by a unitary matrix
$V$
\begin{eqnarray}
{H}_{\rm SF}^{(2)}(m)=
V\pmatrix{&D_{\rm SF}(m)_1\cr D_{\rm SF}(m)_1^\dagger\cr}V^\dagger,\quad
V=\pmatrix{1\cr&\gamma_5u}.
\end{eqnarray}
The determinant of this Dirac operator is evaluated in a single Hilbert
sub-space
\[\det_{\{{\cal H}_-\oplus{\cal H}_-\}}{H}_{\rm SF}^{(2)}(m)\]
and becomes real.

For the other mass term using the physical quark fields of
\eqn{eqn:quark2} and \eqn{eqn:quark3} we also have two flavors Dirac
operator
with
\begin{eqnarray}
D_{\rm SF}^{(2)}(m)=\pmatrix{
\bP_+\left(D+m\eta u\right)\hP_-\cr
&\bP_+\left(D+m\eta u'^\dagger\right)\hP_-},
\end{eqnarray}
which satisfy the same ``Hermiticity'' relation \eqn{eqn:hermiticity2}.

Here we notice that $U(2)$ flavour symmetry is broken to
$U(1)_V\times U(1)_3$ by mass term like chirally twisted Wilson
fermion.
However the symmetry is recovered in massless limit and $m=0$
simulation is possible for relatively small box size we expect that this
flavour symmetry breaking is not a serious problem.

\subsection{$\gamma_5$ mass term}

We have another candidate of the mass term which is consistent with the
orbifolding symmetry \eqn{eqn:symmass}.
That is the $\gamma_5$ mass term.
In order to be able to define the Hermitian Dirac operator we need two
flavors also in this case.
The orbifolded action becomes
\begin{eqnarray}
S=\frac{1}{2}a^4\sum\bpsi D_{\rm SF}(m_5)\psi
\end{eqnarray}
with the SF Dirac operator in the two flavors space
\begin{eqnarray}
D_{\rm SF}(m_5)=
\bP_+\left(D+i\gamma_5\tau^3m_5\left(1-\frac{\ovl{a}}{2}D\right)\right)\hP_-.
\end{eqnarray}
The Hermite conjugate is expressed as
\begin{eqnarray}
D_{\rm SF}(m_5)^\dagger
=\tau^{1,2}\gamma_5uD_{\rm SF}(m_5)u^\dagger\gamma_5\tau^{1,2}
=\tau^{1,2}\gamma_5u'^\dagger D_{\rm SF}(m_5)u'\gamma_5\tau^{1,2}
\end{eqnarray}
and we can easily show that the phase of the determinant is mass
independent in this case.
Furthermore we can absorb the phase into the fields and define the
Hermitian Dirac operator in the same manner
\begin{eqnarray}
H_{\rm SF}(m_5)=D_{\rm SF}(m_5)U^\dagger\gamma_5\tau^{1,2},
\label{eqn:hermite5}
\end{eqnarray}
for which we have exact Hermiticity
$H_{\rm SF}(m_5)^\dagger=H_{\rm SF}(m_5)$
and real determinant.

Up to now the Hermitian Dirac operator \eqn{eqn:hermite0}
\eqn{eqn:hermitem} \eqn{eqn:hermite5} contains gauge dependent unitary
matrix $U^\dagger$, which may be an obstacle for practical application.
This problem may be solved by chirally twisting the fields and adopting
different boundary condition in temporal direction.
We consider following chiral rotation of the fields
\begin{eqnarray}
\psi=e^{-i\frac{\pi}{4}\wh{\gamma}_5\tau^3}\psi',\quad
\bpsi=\bpsi'e^{-i\frac{\pi}{4}\gamma_5\tau^3}.
\end{eqnarray}
The action is re-written as
\begin{eqnarray}
&&
S=\frac{1}{2}a^4\sum\bpsi'\ovl{D}_{\rm SF}(m)\psi',
\\&&
\ovl{D}_{\rm SF}(m)=\wt{\Pi}_-D(m)\wt{\Pi}_-,\quad
D(m)=D+m\left(1-\frac{\ovl{a}}{2}D\right),
\end{eqnarray}
where the projection operator $\wt{\Pi}_\pm$ is defined by using the
time reversal operator $\Sigma$ of \eqn{eqn:timeref0} and third
component of the Pauli matrix
\begin{eqnarray}
\wt{\Pi}_\pm=\frac{1\pm\Sigma\tau^3}{2}.
\end{eqnarray}
We should notice that the new fields $\psi'$ and $\bpsi'$ obey a
different orbifolding condition
\begin{eqnarray}
\wt{\Pi}_+\psi'(x)=0,\quad
\bpsi'(x)\wt{\Pi}_+=0
\end{eqnarray}
and satisfy a different condition at the boundary
\begin{eqnarray}
&&
\left.\wt{P}_+\psi'(x)\right|_{x_0=0}=0,\quad
\left.\wt{P}_-\psi'(x)\right|_{x_0=T}=0,
\\&&
\left.\bpsi'(x)\wt{P}_+\right|_{x_0=0}=0,\quad
\left.\bpsi'(x)\wt{P}_-\right|_{x_0=T}=0
\end{eqnarray}
with a projection
\begin{eqnarray}
\wt{P}_\pm=\frac{1\pm\wt{\Gamma}}{2},\quad
\wt{\Gamma}=i\gamma_5\gamma_0\tau^3.
\end{eqnarray}

Since the boundary condition is completely different, this formalism
may not offer the same SF renormalization scheme.
However it may be possible to define a new renormalization scheme with
this theory.
This formulation gives an equivalent partition function at least for the
massless case connected by the exact chiral symmetry.
This system has a gap caused by the discretization of $p_0$ in
\eqn{eqn:p0} and it may be feasible to make use of the same good
property of the SF formalism in defining the renormalization scale.

It is clear that this Dirac operator gives real determinant with
Hermiticity relation
\begin{eqnarray}
\ovl{D}_{\rm SF}(m)^\dagger
=\gamma_5\tau^{1,2}\ovl{D}_{\rm SF}(m)\gamma_5\tau^{1,2}.
\end{eqnarray}
The flavour symmetry is broken to $U(1)_V\times U(1)_3$ as other massive
theories and $SU(2)_f$ symmetry in massless limit is guaranteed in a
form of ``Ginsparg-Wilson relation''
\begin{eqnarray}
\gamma_5\tau^{1,2}\ovl{D}_{\rm SF}(0)+\ovl{D}_{\rm SF}(0)\gamma_5\tau^{1,2}
=a\ovl{D}_{\rm SF}(0)\gamma_5\tau^{1,2}\ovl{D}_{\rm SF}(0).
\end{eqnarray}

We notice that the projector $\Sigma\tau^3$ commute with the Dirac
operator
\begin{eqnarray}
\left[D(m),\Sigma\tau^3\right]=0
\end{eqnarray}
and the eigenvalue of this new system becomes the same as that of the
ordinary massive overlap Dirac operator $D(m)$ with half numbers of
degeneracy.
The propagator is defined as
\begin{eqnarray}
\ovl{G}_{\rm SF}(x,y)=2\left(\ovl{D}_{\rm SF}(m)^{-1}\right)_{x,y}
=2\left(\wt{\Pi}_-\frac{1}{D(m)}\wt{\Pi}_-\right)_{x,y},
\end{eqnarray}
which takes a simple form at tree level
\begin{eqnarray}
a^3\sum_{\vec{x}}e^{-i\vec{p}\vec{x}}\ovl{G}_{\rm SF}(x,y)&=&
\frac{1}{2aN_T}\sum_{n=-N_T+1}^{N_T}
\frac{-i\gamma_\mu\ovl{A}_\mu+M}{\ovl{A}_\mu^2+M^2}
\nn\\&&\times
e^{ip_0x_0}\left\{
 \left(e^{-ip_0y_0}+e^{ip_0y_0}\right)\wt{P}_-
+\left(e^{-ip_0y_0}-e^{ip_0y_0}\right)\wt{P}_+\right\},
\end{eqnarray}
where
\begin{eqnarray}
&&
\ovl{A}_\mu=\left(1-\frac{\ovl{a}}{2}m\right)A_\mu,\quad
M=m+\left(1-\frac{\ovl{a}}{2}m\right)B
\end{eqnarray}
and $A_\mu$, $B$ are defined in \eqn{eqn:AB}.
The temporal momentum $p_0$ has a discretized value of \eqn{eqn:p0}.

\reseteqnum
\section{Conclusion}
\label{sec:conclusion}

In this paper we propose a new procedure to introduce the SF Dirichlet
boundary condition for general fermion fields.
In the former formulation the boundary condition is introduced by
cutting the Dirac operator at the boundary \cite{Sint94}.
For the Wilson fermion the boundary condition is automatically decided
depending on the signature of the Wilson parameter $r$.
However this formulation produces an exponentially small eigenvalue in
the kernel $D_W$ of the overlap Dirac operator because the relative
signature of the Wilson parameter $r$ and the mass parameter $M$ is
opposite.
This may cause breaking of the locality of the overlap Dirac operator or
breaking of the axial Ward-Takahashi identity in the domain-wall
fermion.

Instead of cutting the Dirac operator we focus on a fact that the chiral
symmetry is broken explicitly in the SF formalism by the boundary
condition and adopt it as a criterion of the procedure.
We also notice that an orbifolded field theory is equivalent to a field
theory with some specific boundary condition.
We search for the orbifolding symmetry which is not consistent with the
chiral symmetry and reproduces the SF Dirichlet boundary condition on
the fixed points.
We found that the orbifolding $S^1/Z_2$ in temporal direction including
the time reflection, the chiral rotation and the anti-periodicity serves
this purpose well.

Application of this procedure to the overlap Dirac operator is
straightforward since this system has both the time reflection and the
chiral symmetry.
The eigenvalue of the Dirac operator and the propagator are derived at
tree level, which are shown to agree with the continuum results in
$a\to0$ limit.

We found a technical problem that the Dirac determinant is complex.
However this is not essential since the phase of the determinant is an
irrelevant ${\cal O}(a)$ term and its variation is localized at the
boundary.
Furthermore we can absorb this phase into re-definition of the fermion
fields for even flavours case and explicit form of Hermitian Dirac
operator was given for two flavours.
For single flavour case it is not still clear how to deal with this
problem.
However the determinant becomes real at tree level and one may expect
that the phase is settled as one approaches to the continuum limit.
Numerical simulation is needed to observe scaling behavior of the
complex phase.

The mass term may still have a problem.
If we define the mass term in an ordinary way to couple to the physical
quark field with kink function $\eta(x_0)$ the phase of the Dirac
determinant becomes mass dependent.
In order to avoid this problem we need even number of flavours, for
which the phase can be absorbed into the fermion field.
The chirally twisted mass term serves for the same purpose.
This mass term is consistent with the orbifolding symmetry and gives
mass independent phase of the determinant, which can also be absorbed by
re-definition of fields.
The most satisfactory solution may be to introduce the chirally twisted
boundary condition, for which the Dirac determinant is naturally real
without any re-definition of fields.
We should notice that the $U(2)$ flavor symmetry is broken to
$U(1)_V\times U(1)_f$ for these two flavors formulation.
However the symmetry is recovered in massless limit and $m=0$
simulation is possible for relatively small box size the flavour
symmetry breaking may not be a serious problem.

The SF formalism with the domain-wall fermion can be formulated in the
same way \cite{YT} since the symmetry is exactly the same as the overlap
Dirac operator \cite{KN}.
Application of this procedure to the Wilson fermion is not trivial since
the Wilson term does not anti-commute with $\Gamma$ operator but commute
with it.
In order to be consistent with the orbifolding symmetry we may need to
introduce the step function $\eta$ of \eqn{eqn:eta} in the Wilson term.
However the Wilson term vanishes on the boundary $x_0=0,T$ and the
doublers appear.
We may need to vanish all the fields at the boundary, with which we
reproduce the same SF Dirac operator as the original one \cite{Sint94}.
Introduction of $\gamma_5$ into the Wilson term may be another way to
apply this procedure.

\section*{Acknowledgement}

I would like to thank M.~L\"uscher for his valuable suggestions and
discussions.
Without his suggestions this work would not have completed.
I also thank to R.~Sommer, S.~Aoki, O.~B\"ar, T.~Izubuchi, Y.~Kikukawa
and Y.~Kuramashi for valuable discussions.
Most of this work was done during my stay at CERN.
I would like to thank a great hospitality of the staffs there.



\begin{thebibliography}{99}

\bibitem{Symanzik}
K.~Symanzik,
Nucl.\ Phys.\ B {\bf 190} (1981) 1.

\bibitem{Luescher85}
M. L\"uscher,
Nucl.\ Phys.\ B {\bf 254} (1985) 52.

\bibitem{RossiTesta}
G.~C.~Rossi and M.~Testa,
Nucl.\ Phys.\ B {\bf 163} (1980)  109; 
Nucl.\ Phys.\ B {\bf 176} (1980) 477; 
Nucl.\ Phys.\ B {\bf 237} (1984) 442; 
Phys.\ Rev.\ D {\bf 29} (1984) 2997; 

\bibitem{Leroy:1990eh}
J.~P.~Leroy, J.~Micheli, G.~C.~Rossi and K.~Yoshida,
Z.\ Phys.\ C {\bf 48} (1990) 653.

\bibitem{LWW}
M. L\"uscher, P. Weisz and U. Wolff,
Nucl.\ Phys.\ B {\bf 359} (1991) 221.

\bibitem{LNWW}
M. L\"uscher, R. Narayanan, P. Weisz and U. Wolff,
Nucl.\ Phys.\ B {\bf 384} (1992) 168
[arXiv:hep-lat/9207009].

\bibitem{Sint94}
S.~Sint,
Nucl.\ Phys.\ B {\bf 421} (1994) 135
[arXiv:hep-lat/9312079].

\bibitem{Sint95} 
S.~Sint,
Nucl.\ Phys.\ B {\bf 451} (1995) 416
[arXiv:hep-lat/9504005].

\bibitem{LSSW9605}
M. L\"uscher, S. Sint, R. Sommer and P. Weisz
Nucl.\ Phys.\ B {\bf 478} (1996) 365
[arXiv:hep-lat/9605038].

\bibitem{LW96}
M.~L\"uscher and P.~Weisz,
Nucl.\ Phys.\ B {\bf 479} (1996) 429
[arXiv:hep-lat/9606016].

\bibitem{Sint01}
S.~Sint,
Nucl.\ Phys.\ Proc.\ Suppl.\  {\bf 94} (2001) 79
[arXiv:hep-lat/0011081].

\bibitem{LSWW93}
M. L\"uscher, R. Sommer, P. Weisz and U. Wolff,
Nucl.\ Phys.\ B {\bf 389} (1993) 247
[arXiv:hep-lat/9207010].

\bibitem{LSWW94}
M. L\"uscher, R. Sommer, P. Weisz and U. Wolff,
Nucl.\ Phys.\ B {\bf 413} (1994) 481
[arXiv:hep-lat/9309005].

\bibitem{alpha95}
G. de Divitiis, R. Frezzotti, M. Guagnelli, M. L\"uscher, R. Petronzio,
R. Sommer, P. Weisz and U. Wolff,
Nucl.\ Phys.\ B {\bf 437} (1995) 447
[arXiv:hep-lat/9411017].

\bibitem{LW95}
M. L\"uscher and P. Weisz,
Phys.\ Lett.\ B {\bf 349} (1995) 165
[arXiv:hep-lat/9502001].

\bibitem{NW}
R.~Narayanan and U.~Wolff,
Nucl.\ Phys.\ B {\bf 444} (1995) 425
[arXiv:hep-lat/9502021].

\bibitem{SS}
S.~Sint and R.~Sommer,
Nucl.\ Phys.\ B {\bf 465} (1996) 71
[arXiv:hep-lat/9508012].

\bibitem{JLLSSSWW}
K. Jansen, C. Liu, M. L\"uscher, H. Simma, S. Sint, R. Sommer,
P. Weisz and U. Wolff,
Phys.\ Lett.\ B {\bf 372} (1996) 275
[arXiv:hep-lat/9512009].

\bibitem{LSSWW}
M. L\"uscher, S. Sint, R. Sommer, P. Weisz and U. Wolff,
Nucl.\ Phys.\ B {\bf 491} (1997) 323
[arXiv:hep-lat/9609035].

\bibitem{LSSW9611}
M.~L\"uscher, S.~Sint, R.~Sommer and H.~Wittig,
Nucl.\ Phys.\ B {\bf 491} (1997) 344
[arXiv:hep-lat/9611015].

\bibitem{JS}
K.~Jansen and R.~Sommer  [ALPHA collaboration],
Nucl.\ Phys.\ B {\bf 530} (1998) 185
[Erratum-ibid.\ B {\bf 643} (2002) 517]
[arXiv:hep-lat/9803017].

\bibitem{CLSW}
S.~Capitani, M.~L\"uscher, R.~Sommer and H.~Wittig  [ALPHA Collaboration],
Nucl.\ Phys.\ B {\bf 544} (1999) 669
[arXiv:hep-lat/9810063].

\bibitem{Neuberger97} 
H.~Neuberger,
Phys.\ Lett.\ B {\bf 417} (1998) 141
[arXiv:hep-lat/9707022].

\bibitem{Neuberger98} 
H.~Neuberger,
Phys.\ Lett.\ B {\bf 427} (1998) 353
[arXiv:hep-lat/9801031].

\bibitem{Kaplan}
D.~B.~Kaplan,
Phys.\ Lett.\ B {\bf 288} (1992) 342
[arXiv:hep-lat/9206013].

\bibitem{Shamir}
Y.~Shamir,
Nucl.\ Phys.\ B {\bf 406} (1993) 90
[arXiv:hep-lat/9303005].

\bibitem{FS}
V.~Furman and Y.~Shamir,
Nucl.\ Phys.\ B {\bf 439}, 54 (1995)
[arXiv:hep-lat/9405004].

\bibitem{KN}
Y.~Kikukawa and T.~Noguchi,
arXiv:hep-lat/9902022.

\bibitem{HJL}
P.~Hernandez, K.~Jansen and M.~L\"uscher,
Nucl.\ Phys.\ B {\bf 552} (1999) 363
[arXiv:hep-lat/9808010].

\bibitem{DHHKN}
T.~DeGrand, A.~Hasenfratz, P.~Hasenfratz, P.~Kunszt and F.~Niedermayer,
Nucl.\ Phys.\ Proc.\ Suppl.\  {\bf 53} (1997) 942
[arXiv:hep-lat/9608056].

\bibitem{BW}
W.~Bietenholz and U.~J.~Wiese,
Nucl.\ Phys.\ B {\bf 464} (1996) 319
[arXiv:hep-lat/9510026].

\bibitem{Hasenfratz}
P.~Hasenfratz,
Nucl.\ Phys.\ B {\bf 525} (1998) 401
[arXiv:hep-lat/9802007].

\bibitem{Luescher98}
M.~Luscher,
Phys.\ Lett.\ B {\bf 428} (1998) 342
[arXiv:hep-lat/9802011].

\bibitem{GW}
P.~H.~Ginsparg and K.~G.~Wilson,
Phys.\ Rev.\ D {\bf 25} (1982) 2649.

\bibitem{YT}
Y. Taniguchi,
in preparation.

\end{thebibliography}
\end{document}